\documentclass[epj]{svjour}
\usepackage{latexsym}
\usepackage{graphics}
\usepackage{amssymb}
\usepackage{amsmath}
\usepackage{graphicx}
\usepackage{bm}
\begin{document}
\title{A Sawtooth Permanent Magnetic Lattice for Ultracold Atoms and BECs}
\author{Amir Mohamadi \and Saeed Ghanbari  
} 
\institute{Department of Physics, Faculty of Science, University of Zanjan, P.O. Box 45195-313, Zanjan, Iran}
\mail{amir.m@znu.ac.ir, sghanbari@znu.ac.ir}
\date{Received: date / Revised version: date}
%
\abstract{We propose a new permanent magnetic lattice for creating periodic arrays of
Ioffe-Pritchard permanent magnetic microtraps for holding and controlling ultracold atoms and
Bose-Einstein condensates (BECs). Lattice can be designed on thin layer of magnetic films
such as $Tb_6$$Gd_10$$Fe_{80}$$Co_4$. In details, we investigate single layer and two crossed layers of
sawtooth magnetic patterns with thicknesses of 50 and 500nm respectively with a
periodicity of 1$\mu$m. Trap depth and frequencies can be changed via an applied bias field to handle
tunneling rates between lattice sites. We present analytical expressions and using numerical calculations
show that this lattice has non-zero
potential minima to avoid majorana spin flips. One advantage of this lattice over previous
ones is that it is easier to manufacture.%
\PACS{
      {67.85.Hj},{67.85.Jk},{37.10.Gh},{37.10.De},{85.85.+j}
     } 
} 
\maketitle
\section{Introduction}
Focused laser beams have been used extensively in years for cooling atoms and molecules
to hold them ~\cite{h.p-02,rmp-98,prl-01,prl-98,Oc-75}. These optical traps are made by a radiation field whose intensity has a
maximum in space. Due to interaction with electric field of laser, small clouds of atoms can
be confined, manipulated and controlled in optical lattices~\cite{epj-05,prl-98-2,arx-428}. Radiation absorption and induced
dipole forces constrain atoms to optical lattice~\cite{p-s-08}. Periodic optical lattices are the result of
interference of intersecting laser beams and have allowed for the observation of the Mott-insulator
to superfluid transition~\cite{pra-10,nat-02}, studies of low-dimensional and quantum degenerate
gases~\cite{prl-06,prl-08} and coherent molecules~\cite{p}. In quantum computation, optical lattices provide
a large qubit system and produce gate operations on the register~\cite{n.o-08,lpr-07}.\\
Using magnetic field is another approach to realize harmonic potential in space~\cite{prl-98-4}.
Magnetic lattices are created using periodically magnetized films or current
carrying wires on atom chips~\cite{jpb-06,jpb-07,prl-99,prl-99-2,epj-05-2}. These magnetic potentials are promising alternatives to the optical
lattices and provide a high degree of design flexibility, arbitrary geometries and lattice
spacing. Magnetic lattices can produce highly stable periodic potential wells with low noise~\cite{rmp-07}. Single 3D magnetic microtraps also have been introduced either by current carrying wires or
permanent magnetic slabs~\cite{pra-09,arx-10}. Recently, development of technology has let us for working
on high-quality magnetic materials based on permanent, perpendicularly magnetized films~\cite{pra-08,pra-07}. Periodic grooved magnetic mirrors have been made by the technology of thin films to
reflect ultracold atoms~\cite{pra-09-2}. In the presence of magnetic bias fields, mirrors are transferred into
1D arrays of 2D waveguides and 2D arrays of 3D traps~\cite{jpb-06}. Magnetic films such as
$Tb_6$$Gd_10$$Fe_{80}$$Co_4$ have excellent magnetic properties; homogeneity, high magnetization, large coercivity ($H_{\rm c}\sim$3k$O_{\rm e}$) and high Curie temperature($\sim$300$^{o}$C)~\cite{jpb-06,jpb-07}.\\
In this paper, we introduce a new atom chip based on permanent periodic grooved magnetic
film and external magnetic fields. This sawtooth magnetic lattice is proposed for
controlling, manipulating and trapping clouds of ultracold atoms when they are in low-field
seeking states. Analytical expressions and numerical calculations show that the magnetic field minima are non-zero.  Therefore, ultracold atoms
can be hold and trapped without any spin flips in lattice sites. In particular, we investigate 1D
array of this grooved magnetic lattice, when bias fields are applied. Results show that
produced waveguides have controllable depth and frequencies to handle clouds of quantum
degenerate gases on the chip surface. We show that two crossed separated layers of periodic
arrays also can produce 2D magnetic lattice of 3D microtraps when bias fields are applied.
The microtraps have non-zero minimum and high depth ($\sim$0.2mK) which cause
prevention of majorana spin flips and efficient restriction of trapped atoms, respectively.
Magnetic atom chips based on permanent magnetic materials have some advantages over
optical or microwire devices. Since there is no spontaneous emission due to decay of excited
atoms, light scattering and decoherence of cold atoms vanishes and no beam alignment is
required~\cite{p-s-08}. Heat dissipation and loss of atoms due to fluctuation of current in current-carrying
microwires are prevented in permanent magnetic lattices. They can produce large field
curvatures, large trap depth and high frequencies with very low technical noises~\cite{rmp-07}. the Main
motivation of this work is introducing a more flexible permanent magnetic lattice compared with the configuration of the two crossed arrays of rectangular permanent magnetic slabs proposed in~\cite{jpb-06}. Considering the shape of this new atom chip, current technology in grooving of films for magnetic
mirrors allows us to produce it more easily compared with previous ones.
\begin{figure}
\begin{center}
\includegraphics[width=8.0cm]{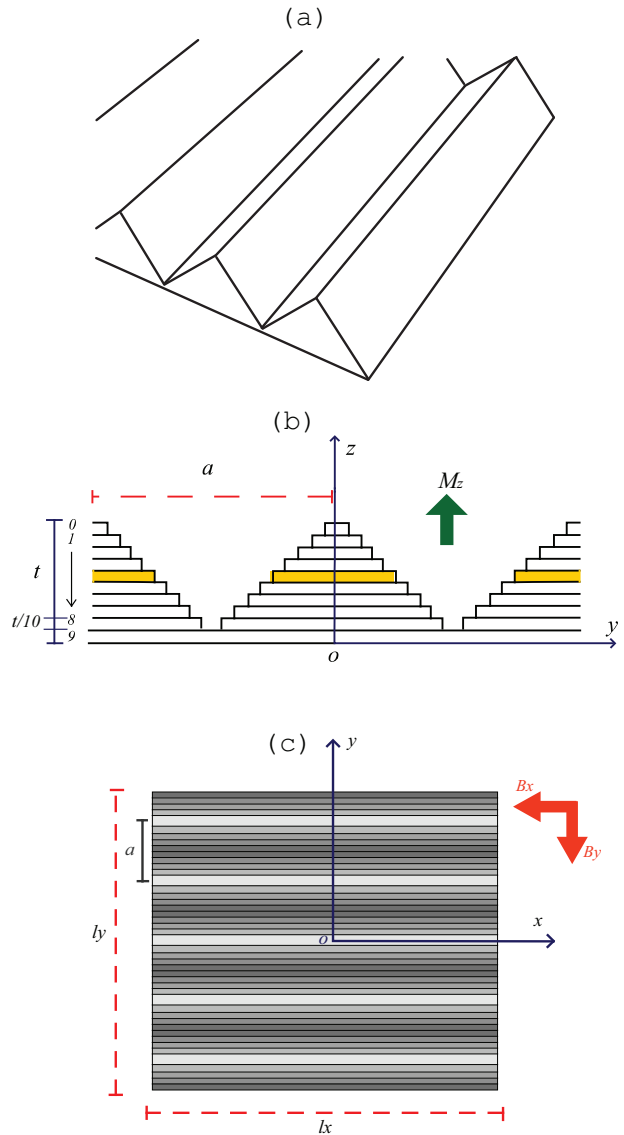}
\caption{(Color online)(a) 1D array of sawtooth permanent magnetic lattice.  (b)  Schematic of sawtooth permanent magnetic lattice film which has been estimated by $n$=10 sub-layers of rectangular magnetic slabs. Periodicity of the sawtooth permanent magnetic lattice is the periodicity of the rectangular slabs and is defined by distance of two alternative peaks of the structure.  Lattice has the periodicity $a$=1$\mu$m, magnetization $M_z$=3.8 kG along the $z$-direction and thickness $t$=500nm.  For $n$=10, thickness of each sublayer is $t/n$=50nm while width of slabs is changed according to $w_i$= $a$($i$+1)/$n$.  (c)   $l_x$ and $l_y$, dimensions of the atom chip in $x$- and $y$- directions, respectively.  For $n_r$=1000 sawtooth magnetic slabs $l_x$= $l_y$=1mm.  External magnetic field is applied along $x$- and $y$-directions to produce waveguides with non-zero potential minima.} \label{fig1}
\end{center}
\end{figure}

\section{Analytical expressions for Magnetic lattices}
We study first the simple case of a single layer of sawtooth periodic array of grooved magnetic films with a thickness of $t$, periodicity $a$ along the $y$-direction, magnetization $M$=$M_z$$\widehat{z}$  and constant bias magnetic fields $B_{1x}$, $B_{1y}$ and $B_{1z}$ along the $x$-, $y$- and
$z$-directions.
\subsection{Atom waveguide}
The configuration of the permanent magnets is shown in Fig.1. Pattern of magnetic film has a
periodicity (distance between two alternative peaks) of $a=1\mu$m and a thickness of $t=500$nm while
dimensions of lattice in $x$- and $y$- directions, $l_x$ and $l_y$ are 1mm (Fig.1(c)). For analytical and
numerical calculations, we approximate our sawtooth magnetic pattern thin rectangular magnetic slabs same magnetization and
periodicity but with different widths. As Fig.1 (a) shows, infinite periodic arrays of magnets are placed above each
other to shape sawtooth structure. At a large distance where $k$($z-t$)$\gg$1, the using expressions given in Ref.~\cite{jpb-06},
the components of the magnetic field can be written

 \begin{eqnarray}
  B_{\rm x}&=& B_{\rm 1x}\\
  B_{\rm y}&=&B_{\rm 0\gamma}\sin{(ky)}e ^{-kz}+B_{\rm 1y}\nonumber \\
 B_{\rm z}&=&B_{\rm 0\gamma}\cos{(ky)}e ^{-kz}+B_{\rm 1z}\nonumber
  \end{eqnarray}

where $k$=2$\pi/a$ and $B_{\rm 0\gamma}=B_{\rm 0}\sum_{i=0}^{n-1}\gamma_{\rm \it i}$, respectively.  $B_{\rm 0}$ and  $\gamma_{\rm \it i}$ are $M_{\rm z}$/$\pi$(Gaussian units) and related coefficient of the $\it i$th infinite periodic magnets on atom chip [see Fig.1(b)], respectively. $\gamma_{\rm \it i}$ is determined from the series [for more details, see section 2.3]
\begin{eqnarray}
\gamma_{\rm \it i}&=&\frac{1}{A}\sum_{j=0}^{i}\cos{(\frac{(2j+1)ka}{4n})}(e^{k\frac{t}{n}}-1)e^{kt(1-\frac{(i+1)}{n})} \\ \nonumber
A&=&\sum_{j=0}^{n/2-1}\cos{(\frac{(2j+1)ka}{4n})}
\end{eqnarray}

where $n$ is the number of sub-layers (i.e. order of approximation).  The magnitude of the magnetic field above surface can be written

 \begin{eqnarray}
 B(y,z)&=& (B_{\rm 1x}^2+B_{\rm 1y}^2+B_{\rm 1z}^2+ 2[B_{\rm 0\gamma}B_{\rm 1y}\sin{(ky)} +\\&B&_{\rm 0\gamma}B_{\rm 1z}\cos{(ky)}]e^{-kz}+ B_{\rm 0\gamma}^{2}e^{-2kz})^{1/2}\nonumber
 \\ \nonumber
\end{eqnarray}%

In simple case when we take $B_{\rm 1z}$=0, the minimum of magnetic field in 1D array of periodic magnetic lattice of 2D microtraps is
\begin{equation}
B_{\rm min}=|B_{\rm 1x}|
\end{equation}%
where the center of the waveguides are located at
\begin{eqnarray}
y_{\rm min}&=&(n_{\rm y}+\frac{1}{4})a
\\ z_{\rm min}&=&\frac{1}{k}\ln{(\frac{B_{\rm 0\gamma}}{|B_{\rm 1y}|})}\nonumber
\end{eqnarray}%
where $n_{\rm y}$ is the trap number in the $y$-direction and takes values of 0, $\pm$ 1, $\pm$ 2,... .\\
According to equation 4, the $x$ component of the bias field is necessary to avoid spin flips near the center of 2D microtraps.
The magnetic barrier heights along $y$- and $z$-directions which depend on the bias fields are given by
\begin{eqnarray}
\Delta B^{y}&=&B^{y}_{\rm max}-B_{\rm min}=(B_{\rm 1x}^{2}+4B_{\rm 1y}^{2})^{1/2}-|B_{\rm 1x}|\\
\Delta B^{z}&=&B^{z}_{\rm max}-B_{\rm min}=(B_{\rm 1x}^{2}+B_{\rm 1y}^{2})^{1/2}-|B_{\rm 1x}|\nonumber
\end{eqnarray}%

Second derivatives (curvatures) of the magnetic field at the minimum of the potential and trap frequencies of the waveguides along $y$- and $z$-directions could be analytically determined by

\begin{eqnarray}
\frac{\partial^{2}B}{\partial y^{2}}&=&\frac{\partial^{2}B}{\partial z^{2}}=k^2\frac{B_{\rm 1y}^2}{|B_{\rm 1x}|}
\\\omega_{\rm y}&=&\omega_{\rm z}=kB_{\rm 1y} (\frac{\mu_{\rm_{\rm B}}g_{\rm_{\rm F}}m_{\rm_{\rm F}}}{mB_{\rm 1x}})^{1/2}
\end{eqnarray}

where $\mu_{\rm_{\rm B}}$,
$g_{\rm_{\rm F}}$, $m_{\rm_{\rm F}}$ and $\it m$ are the Bohr magneton, Lande factor, magnetic quantum number and mass of atom, respectively.
Equations (5) and (8) show that increasing $B_{\rm 1y}$ reduces $z_{\rm min}$ and increases frequencies of the traps, respectively.
The frequencies of traps are independent of magnetization of the magnetic film.  They depend on the external magnetic fields and the periodicity of magnetic lattice.

\subsection{2D permanent magnetic lattice of microtraps}
\begin{figure}
\begin{center}
\includegraphics[width=8.0cm]{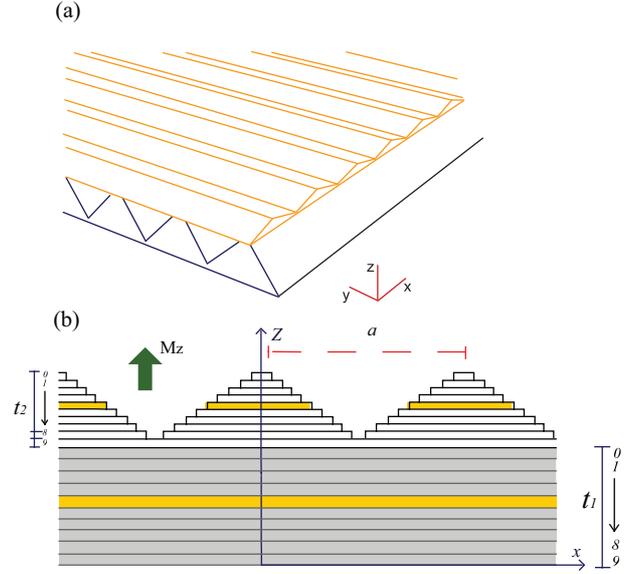}
\caption{(Color online)(a) Two layers of sawtooth configuration permanent magnetic lattice .  (b) Schematic of (a) estimated by $n$=10 sub-layers of rectangular magnetic slabs for each layer. Top and bottom arrays have corresponding coefficients series $\lambda_i$ and  $\gamma_i$, respectively in magnetic field.  Periodicity of top and bottom arrays are $a$=1$\mu$m along both $x$- and $y$-directions.  Arrays have magnetization $M_z$=3.8 kG along the $z$-direction and thickness $t_1$=500nm and $t_2$=50nm for bottom and top arrays, respectively.} \label{fig2}
\end{center}
\end{figure}

In order to produce Ioffe-Pritchard microtraps, two  crossed layers of periodic arrays of sawtooth magnetic films with two different thicknesses of $t_1$ and $t_2$, separation of $d$ and same magnetization $M_z$ with uniform bias fields is investigated (Fig.2).  The top and bottom arrays have same periodicity $a$ along $x$- and $y$-directions.  Thickness of bottom array, $t_1$ is larger than thickness of top array, $t_2$.  $M_z$ is along z-direction and bias fields are $B_{1x}$ and $B_{1y}$.
The components of the magnetic field, for the large distances (k(z-$t_1$-$t_2$)$\gg$1) from top of the surface can be obtained by
\begin{eqnarray}
 B_{\rm x}&=& B_{\rm 0\lambda}\sin{(kx)}e ^{-kz}+B_{\rm 1x}
 \\ B_{\rm y}&=&B_{\rm 0\gamma}\sin{(ky)}e ^{-kz}+B_{\rm 1y}\nonumber
 \\ B_{\rm z}&=&(B_{\rm 0\gamma}\cos{(ky)}+B_{\rm 0\lambda}\cos{(kx)})e ^{-kz}+B_{\rm 1z}\nonumber
\end{eqnarray}%
where $B_{\rm 0\lambda}=B_{\rm 0}\sum\lambda_{\rm \it i}$.  Coefficient of the $\it i$th sublayer of infinite magnets of the top array, $\lambda_{\rm \it i}$ is determined, similary

\begin{eqnarray}
\nonumber\lambda_{\rm \it i}=\frac{1}{A}\sum_{j=0}^{i}\cos{(\frac{(2j+1)ka}{4n})}(e^{k\frac{t_{\rm 2}}{n}}-1)e^{k((1-\frac{(i+1)}{n})t_{\rm 2}+t_{\rm 1}+d)}\\
\end{eqnarray}

where $\it d$ is the gap between bottom and top arrays.  The magnitude of magnetic field above the surface can be written

 \begin{eqnarray}
\nonumber B(x,y,z)= (B_{\rm 1x}^2+B_{\rm 1y}^2+B_{\rm 1z}^2
 + 2[B_{\rm 0\lambda}B_{\rm 1x}\sin{(kx)} \\ \nonumber
 +B_{\rm 0\gamma}B_{\rm 1y}\sin{(ky)}+B_{\rm 0\lambda}B_{\rm z}\cos{(kx}) \\  \nonumber
 +B_{\rm 0\gamma}B_{\rm z}\cos{(ky)}]e^{-kz}+[B_{\rm 0\lambda}^2+B_{\rm 0\gamma}^2\\
 +2B_{\rm 0\lambda}B_{\rm 0\gamma}\cos{(kx)}\cos{(ky)}]e^{-2kz})^{1/2}
\end{eqnarray}%
Non-zero minimum of the magnetic field and the center of microtraps in $x$-,$y$- and z-directions for $B_{1z}$=0 can be determined by

 \begin{eqnarray}
 B_{\rm min}&=& \frac{|B_{\rm 1x}B_{\rm 0\lambda}-B_{\rm 1y}B_{\rm 0\gamma}|}{(B_{\rm 0\gamma}^2+B_{\rm 0\lambda}^2)^{1/2}}\\
 x_{\rm min}&=&(n_{\rm x}+\frac{1}{4})a \\ \nonumber
 y_{\rm min}&=&(n_{\rm y}+\frac{1}{4})a \\ \nonumber
 z_{\rm min}&=&\frac{1}{k}\ln{(\frac{\alpha B_{\rm 0\lambda}}{B_{\rm 1x}})} \nonumber
 \end{eqnarray}%

where $n_{\rm x}$ and $n_{\rm y}$ are number of microtraps and take 0, $\pm$1, $\pm$2,... values.  Dimensionless constant $\alpha$ is (1/2)(1+1/$\alpha_{\rm 0}^2)$ where $\alpha_{\rm 0}=B_{\rm 0\lambda}/B_{\rm 0\gamma}$.  At the center of 3D microtraps, the curvature of magnetic fields and barrier heights can be obtained by
\begin{eqnarray}
\frac{\partial^{2}B}{\partial x^{2}}&=&\frac{\partial^{2}B}{\partial y^{2}}=\frac{1}{2}\frac{\partial^{2}B}{\partial z^{2}}=k^2\alpha_{1}|B_{\rm 1x}| \\
\Delta B^{x}&=&\Delta B^{y}=\alpha_{2}|B_{\rm 1x}|\\ \nonumber
\Delta B^{z}&=&(1+\alpha_{0}^{2})^{1/2}|B_{1x}|
\end{eqnarray}%
where $\alpha_{1}$ and $\alpha_{2}$ are
 \begin{eqnarray}
 \alpha_1&=&\frac{2\alpha_{0}^2}{(1+\alpha_{0}^2)^{1/2} |1-\alpha_0|} \\ \nonumber
  \alpha_2&=&(1+\alpha_0^{2}+\frac{4\alpha_{0}^2}{1+\alpha_{0}^2})^{1/2}-\frac{|1-\alpha_0^2|}{(1+\alpha_0^2)^{1/2}}
 \end{eqnarray}
The frequencies of microtraps in three directions are

 \begin{eqnarray}
 \omega_{\rm x}=\omega_{\rm y}=\frac{\omega_{\rm z}}{\sqrt{2}}=k(\frac{\mu_{\rm_{\rm B}}g_{\rm_{\rm F}}m_{\rm_{\rm F}}B_{\rm 1x}\alpha_{1}}{m})^{1/2}
\end{eqnarray}%

For a symmetrical 2D magnetic lattice, same magnetic height barriers in $x$- and $y$-directions, $\Delta B^{x}=\Delta B^{y}$ are needed.  The constraint of  $B_{1y}=\alpha_0 B_{1x}$ satisfies our request where $\alpha_0=B_{0\lambda}/B_{0\gamma}$.
\subsection{Corresponding magnetic coefficients of estimated layers}
\begin{figure}
\begin{center}
\includegraphics[width=6.0cm]{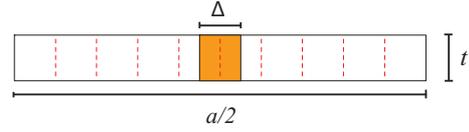}
\caption{ (Color online) Schematic of one slab in permanent magnetic array with periodicity $a$ and width $a$/2.  Slabs divided into arbitrary (e.g. $n$=10) narrow slabs $\Delta$ with same periodicity.  The magnetic field of narrow slabs related to initial one by coefficient $c_1$.  In Eq.19 to obtain $c_1$, slab $\Delta$ shifted to right and left sides by small steps $\Delta$($i$+1/2) , where $i$=0, 1,$...$, $n$/2-1.  Summations are taken on all dashed spaces and results are compared with magnetic field of large slab to find $c_1$.} \label{fig3}
\end{center}
\end{figure}
The components of magnetic field for $n_r$ infinitely long and thin permanent magnetic slabs along the $x$-direction arranged with periodicity $a$ along the $y$- direction and separation distance $a$/2 between the center of neighboring slabs, are given by~\cite{apb-02}
 \begin{eqnarray}
   \\ B_{\rm x}&=&0\nonumber
  \\ B_{\rm y}&=&B_{\rm 0y}\sin{(ky)}e ^{-kz}\nonumber
 \\ B_{\rm z}&=&B_{\rm 0y}\cos{(ky)}e ^{-kz}\nonumber
\end{eqnarray}%
where $k$ and  $B_{0y}$  are  2$\pi$/$a$ and $B_0$(e$^{kt}$-1), respectively.  As we mentioned earlier in 2.1 and 2.2, permanent magnetic slabs with sawtooth geometry can be estimated by $n$ multi-layer arrays of rectangular magnets with same periodicity $a$ and different width $w_i$, where $i$ is the number of a specific sublayer.  The major problem is the calculation of magnetic field of sub-layers with $w_i>$$a$/2 or $w_i<$$a$/2.  In other words, for magnetic slabs with different width and separation, Eq.18 does not work and needs to be modified.   As Fig.1 (b) shows, width plus separation between neighboring slabs for each sublayer $i$ is constant and equal to periodicity of the lattice.  It means that for any modification of Eq.18, $k$ remains unchanged.  The new components of magnetic field for slabs with arbitrary width are related to Eq.18 by coefficient $\gamma$.  The slabs of periodic lattice by width and separated gap $a$/2 can be divided into $n$ narrow slabs (Fig.3) with width $\Delta$=$a$/2$n$.  The magnitude of magnetic field due to these narrow slabs is related to the initial slab with coefficient $c_1$.  For the calculating of $c_1$, narrow slabs can be shifted by small steps of $a$/4$n$ along $y$ axis to make initial array.  The component of the magnetic field for these shifted narrow slabs, for example in $y$-direction must be consistent with Eq.18, so we have
\begin{eqnarray}
B_{\Delta,y}&=&c_1B_{0y}\sum_{j=0}^{n/2-1}\sin{k(y\pm(2j+1)\Delta)}e^{-kz}\\ \nonumber
&=&B_{\rm 0y}\sin{(ky)}e ^{-kz}
\end{eqnarray}
where $c_1$ is determined
\begin{eqnarray}
c_1=(2\sum_{j=0}^{n/2-1}\cos{k(y+(2j+1)k\Delta)})^{-1}
\end{eqnarray}
We use these narrow slabs $\Delta$ as unit to design any periodic lattice by arbitrary slab width $w_i$ and periodicity $a$. In general, the $\Delta$ slabs must be several times shifted to the right and left to model the given periodic array.  These shifts lead to another series in calculation of the magnetic field which is given by

\begin{eqnarray}
\gamma_i=\frac{2}{c_1}\sum_{j=0}^{i}\cos{(2j+1)k\Delta}(e^{k\frac{t}{n}}-1)e^{kt(1-\frac{i+1}{n}})
\end{eqnarray}
For sawtooth array, all estimated sub-layers have same thickness $t/n$ and distance from origin $s_i$= 1-($i$+1)/$n$. Summation on $\gamma_i$ and $\lambda_i$ for $n$=10 sub-layers are determined 13.97 and 5.47, while for $n$=20 sub-layers reduce to 13.02 and 5.45 values, respectively.
\section{Numerical calculations for magnetic lattice}
Numerical calculations for finite arrays of sawtooth magnetic films with finite length were made.  Radia Package~\cite{rad}, interfaced to Mathematica software was used to numerically obtain magnetic field for different configurations of magnets in 1D array of waveguides and 2D arrays of 3D microtraps.
\subsection{Atom waveguide}
Numerical calculations of 1D array of waveguides above the sawtooth surface of the atom chip were performed when bias magnetic field along the $y$-direction, $B_{1y}$ was applied.  Thickness $t$ =500nm, periodicity $a$=1$\mu$m, magnetization 4$\pi$$M_z$=3.8kG, finite toothy magnets number $n_r$=5001 which is approximated by $n$=10 rectangular slabs.  In Fig.4, the components of magnetic field of atom chip above the surface are shown where $B_{1y}$= $-$5G.  This configuration may be used as spatial grating to diffract slowly moving atoms.  By including $B_{1x}$= $-$3G parallel to the waveguide axis, center of waveguides becomes non-zero with $B_{min}$=$|B_{1x}|$=3G to avoid spin flips of cold atoms.  Center of waveguides is at $z_{min} $=1.29$\mu$m from bottom of the layer.  Magnetic barrier heights along $y$- and $z$-directions are 7.4 and 2.8G, respectively.  The trap frequencies of 2D microtrap in $y$- and $z$-directions are numerically determined 2$\pi\times$36.7 kHz for $^{87}$Rb atoms.  Fig.4 (a) and (b) show excellent agreement with numerical calculations and analytical expressions of magnetic field along $y$- and $z$-directions for $z-t\gg$$a$/2$\pi$.   Numerically calculated parameters and related analytical expressions for a sawtooth magnetic film with bias fields are listed in table 1. \\
\begin{table*}
\begin{center}
 \caption{\label{table1}.  Numerical and analytical determinations for single array of sawtooth permanent magnetic lattice.  Lattice has periodicity $a$=1$\mu$m along $y$-direction, thickness $t$=500nm and magnetization $M_z$=3.8 kG along $z$-direction (Fig.1(b)).  Results is obtained for $^{87}$Rb atoms in different approximation orders (sub-layers) with bias field $B_{1x}$=$-$3G and $B_{1y}$=$-$5G.}
\hspace{-1.2cm}
\begin{tabular}{llcc}
\hline

Parameter&Definition&\hspace{.25cm}Numerical&\hspace{.25cm}Analytical \\
&&$n$=10\hspace{.5cm}$n$=20&$n$=10\hspace{.5cm}$n$=20\\
\hline

\verb   $n_r$    & Number of magnets in $y$-direction & 5001              & $\infty$     \vspace{.1cm}            \\
\verb   $y_{min}$($\mu$m)& Center of  first waveguide in $y$-direction & 0.25              &0.25 \vspace{.1cm}                \\
\verb   $z_{min}$($\mu$m)& Center of  waveguides  in $z$-direction          &1.29\hspace{.5cm}1.28              &1.29\hspace{.5cm}1.28     \vspace{.1cm}         \\
\verb  $\frac{\partial^2 B}{\partial y^2}\hspace{.1cm}({G\over{cm}^2})$  & Curvature of  B in $y$-direction  &   3.29$\times$10$^{10}$ &   3.28$\times$10$^{10}$  \vspace{.1cm}   \\
\verb   $\frac{\partial^2 B}{\partial z^2}\hspace{.1cm}({G\over{cm}^2})$             & Curvature of  B in $z$-direction  &3.29$\times$10$^{10}$&   3.28$\times$10$^{10}$  \vspace{.1cm}\\
\verb  $\omega_y$/2$\pi$(kHz)             & Trap frequency in  $y$- direction &   231.5&    231.5 \vspace{.1cm}  \vspace{.1cm} \\
\verb  $\omega_z$/2$\pi$(kHz)             & Trap frequency in  $z$- direction &   231.5 &   231.5 \vspace{.1cm}   \vspace{.1cm} \\
\verb  $ B_{min}$ (G)           & Minimum of B at center of waveguides & 3& 3\\
\end{tabular}
\end{center}
\end{table*}
\begin{table*}
\begin{center}
 \caption{\label{table2}Numerical and analytical determinations for two crossed layers of sawtooth permanent magnetic lattice.  Lattice has periodicity $a$=1$\mu$m along $x$- and $y$-directions, with bottom and top thicknesses  $t_1$=500nm and $t_2$=50nm respectively. Two layers have same magnetization $M_z$=3.8 kG along $z$-direction (Fig.2(b)).  Results are obtained for $^{87}$Rb atoms and bias field $B_{1x}$=$ -$5.6G and $B_{1y}$= $-$2.18G for $n$=10 sublayer and $B_{1x}$=$ -$5.6G and $B_{1y}$=$ -$2.3G for $n$=20 sublayer.}
\hspace{-1.2cm}
\begin{tabular}{clcc}
\hline

Parameter&Definition&\hspace{.25cm}Numerical&\hspace{.25cm}Analytical \\
&&$n$=10\hspace{.5cm}$n$=20&$n$=10\hspace{.5cm}$n$=20\\
\hline

\verb  $n_r$    & Number of magnets in $x$- and $y$-direction         & 5001              &$\infty$        \vspace{.1cm}         \\

\verb   $x_{min}$($\mu$m)& Center of  first microtrap in $x$-direction          &0.25              & 0.25 \vspace{.1cm}               \\

\verb   $y_{min}$($\mu$m)& Center of  first microtrap in $y$-direction             & 0.25              &0.25  \vspace{.1cm}              \\
\
\verb   $z_{min}$($\mu$m)& Center of  microtrap in $z$-direction              & 1.33\hspace{.5cm}1.31              &1.33\hspace{.5cm}1.31 \vspace{.1cm}\\

\verb  $\frac{\partial^2 B}{\partial x^2}\hspace{.1cm}({G\over{cm}^2})$ & Curvature of  B in  $x$-direction  & 2.98$\times$10$^{10}$ &  2.98$\times$10$^{10}$ \vspace{.1cm}    \\

\verb  $\frac{\partial^2 B}{\partial y^2}\hspace{.1cm}({G\over{cm}^2})$& Curvature of  B in  $y$-direction   &2.98$\times$10$^{10}$ & 2.98$\times$10$^{10}$ \vspace{.1cm} \\

\verb  $\frac{\partial^2 B}{\partial z^2}\hspace{.1cm}({G\over{cm}^2})$ & Curvature of B in  $z$-direction  & 5.95$\times$10$^{10}$\hspace{.3cm}5.97$\times$10$^{10}$ &   5.96$\times$10$^{10}$\hspace{.3cm}5.97$\times$10$^{10}$ \vspace{.1cm}   \\

\verb  $\omega_x$/2$\pi$(kHz)             & Trap frequency in  $x$- direction & 220\hspace{.5cm}220.5&220.3\hspace{.5cm}220.5 \vspace{.1cm}\\
\verb  $\omega_y$/2$\pi$(kHz)             & Trap frequency in  $y$- direction &220\hspace{.5cm}220.5& 220.3\hspace{.5cm}220.5 \vspace{.1cm}\\
\verb  $\omega_z$/2$\pi$(kHz)             & Trap frequency in  $z$- direction &311.6\hspace{.5cm}312& 311.6\hspace{.5cm}312 \vspace{.1cm} \\
\verb  $ B_{min}$   (G)         & Minimum of B at center of microtraps  &  4.41\hspace{.5cm}4.27& 4.41\hspace{.5cm}4.27 \\

\end{tabular}
\end{center}
\end{table*}
 Large curvatures of the magnetic field ($\sim$10$^{10}$ G/cm$^2$) can be realized in the permanent magnetic lattices due to their high magnetization which a small size of the microtrap is led to high frequencies for efficient trapping and controlling of ultracold atoms.  Rate of spin flips transition for single atom in ground state of harmonic waveguide, when nuclear spin is neglected and total spin is spin of single electron, can be written~\cite{rmp-07}
\begin{figure}
\begin{center}
\includegraphics[width=8.0cm]{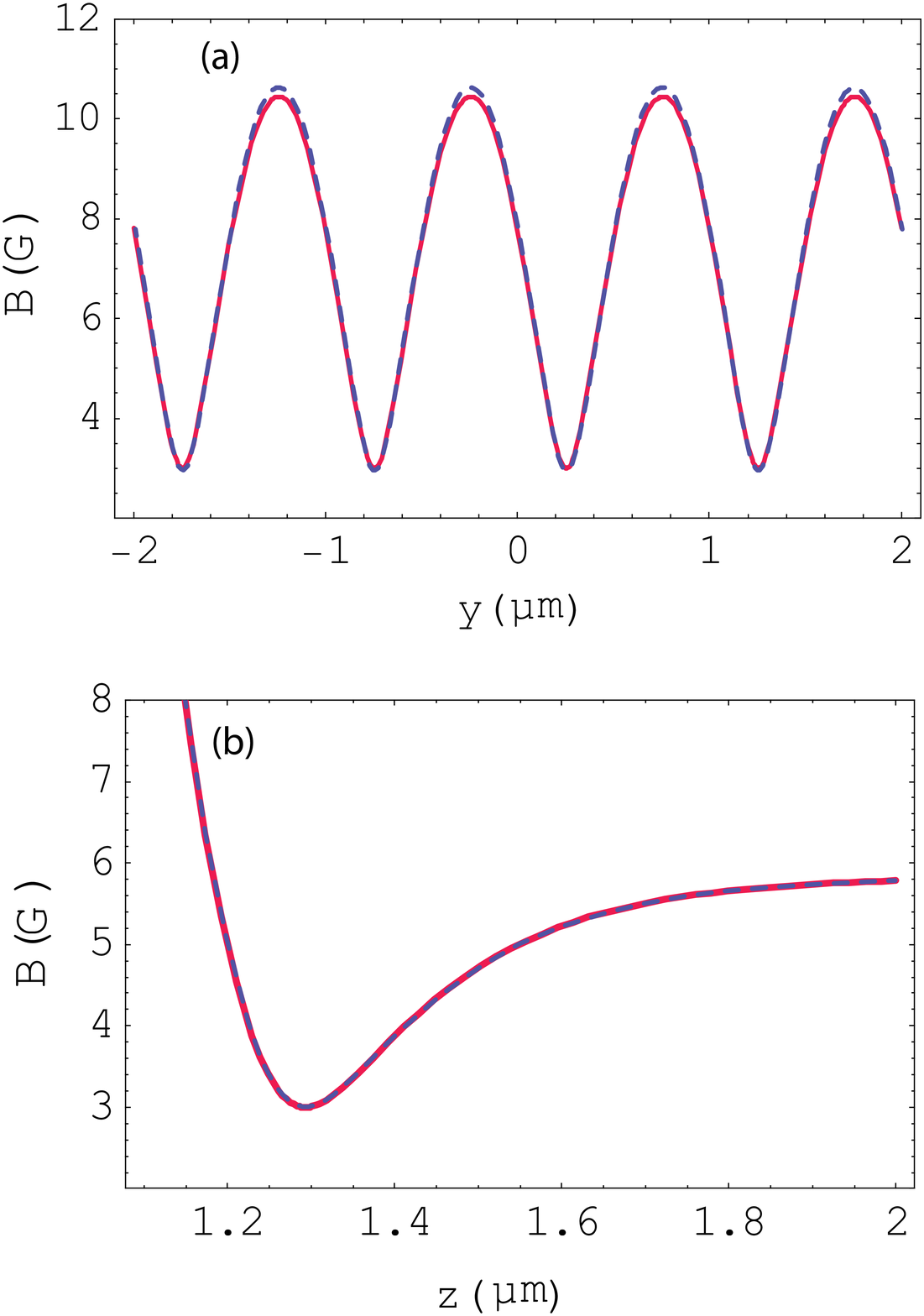}
\caption{(Color online) (a) and (b) The magnitude of magnetic field close to the center of the waveguides in $y$-and $z$-directions, respectively.  Numerical calculations (red solid) and analytical expressions (dashed-blue) are obtained using Radia and Eq.3, respectively, for $n$=10 estimated sub-layers.  Plots show good versatility between numerical and analytical determinations for $n_r$=5001 sawtooth slabs with magnetic bias field $B_{min}$=$|B_{1x}|$= 3G and $B_{1y}$=$-$5G.  Center of the waveguides is at $z_{min}$=1.29$\mu$m and $y_{min}$=($n_y$+1/4)$a$, where $n_y$ is number of waveguides.} \label{fig4}
\end{center}
\end{figure}
\begin{figure*}
\begin{center}
\includegraphics[width=17.0cm]{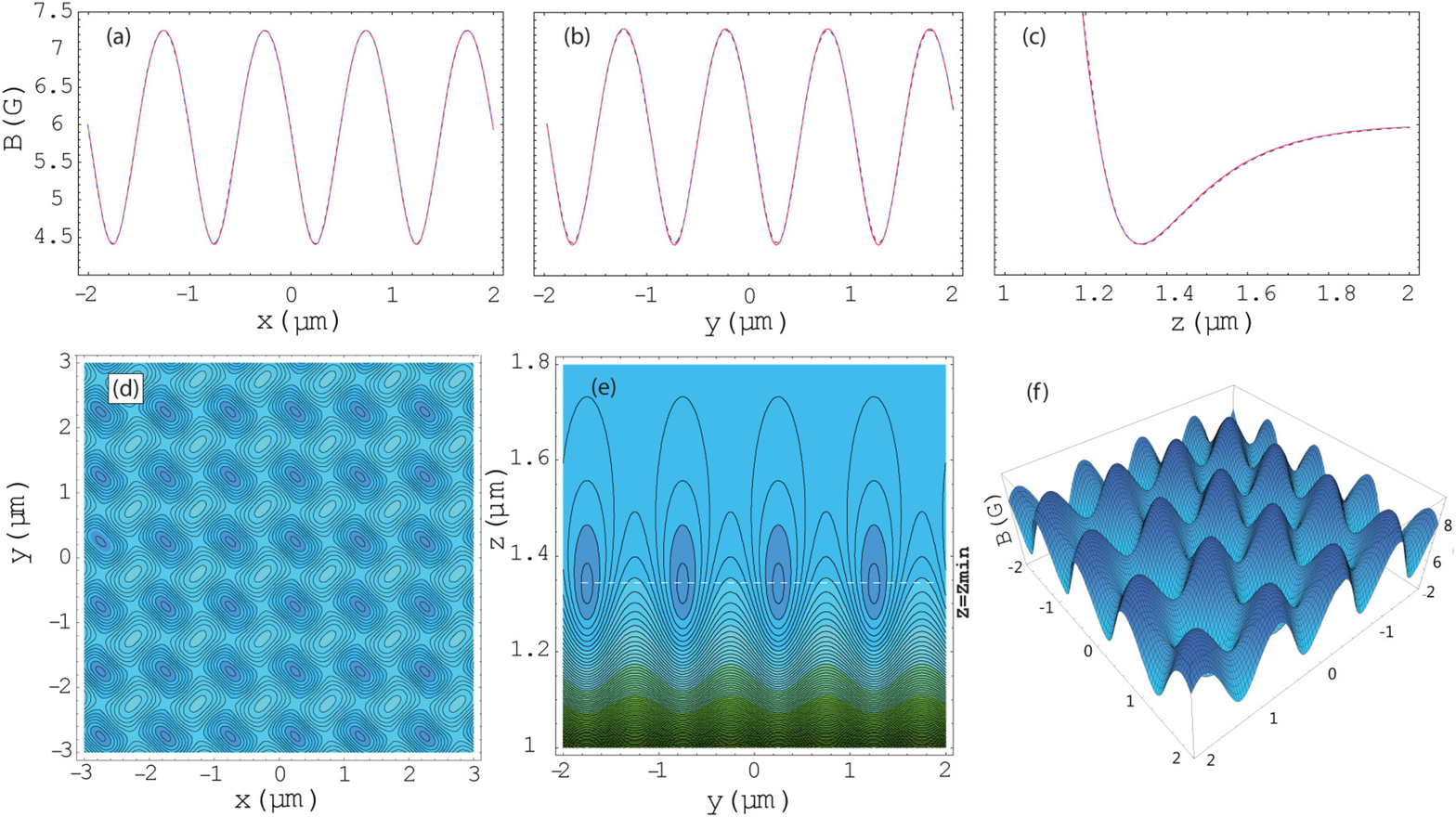}
\caption{(Color online)(a), (b) and (c) The magnitude of magnetic field close to the center of the microtraps in $x$-, $y$- and $z$-directions, respectively.  Numerical calculations (red solid) and analytical expressions (dashed-blue) obtained by Radia and Eq.11 respectively, for $n$=10 estimated layers .  Plots show excellent versatility between numerical and analytical determinations for $n_r$=5001 sawtooth slabs with magnetic bias field $B_{1x}$=$-$5.6G, $B_{1y}$=$-$2.18G and $B_{min}$=4.41G.  Center of microtraps along $z$-direction is at $z_{min}$=1.33$\mu$m.  Position of first microtrap along $x$-and $y$-direction is $x_{min}$=$y_{min}$=0.25 $\mu$m which is repeated by periodicity of arrays.  (d) and(e) Contour plots of magnetic field at $z$=$z_{min}$ and $x$=$x_{min}$ planes, respectively.  Yellow points (grey, in black and white) show positions of microtraps .  The density of lattice sites is $\sim$10$^{6}$ traps/mm$^{2}$. (f) 3D plots of magnetic field on the surface of atom chip at $z$=$z_{min}$ plane.   } \label{fig5}
\end{center}
\end{figure*}
\begin{eqnarray}
\Gamma=\frac{\pi\omega}{2}e^{(\frac{-U_{min}}{\hbar\omega}-\frac{1}{2})}
\end{eqnarray}
where $U_{min}$ and $\omega$ are potential energy minimum and radial oscillation frequency, respectively.  Eq.22 works for $U_{min}\gg\hbar\omega$.  By using table 1, where $U_{min}$/$k_B$=201.5$\mu$k and $\hbar\omega$/$k_B$=1.7$\mu$k, decay life time of trapped $^{87}$Rb atoms in F=2 and $m_F$=2 state are so long and there are no majorana losses in the  waveguides.  Thereby, the introduced waveguides can be used to transfer cold atoms over the surface of single layer atom chip along $x$-direction without any significat atom losses from the harmonic trap.
\subsection{2D permanent magnetic lattice of microtraps}
Here, we investigate two crossed sawtooth magnetic layers to create 3D Ioffe-Pritchard microtraps.  This magnetic lattice is created when bias fields are applied.  Fig.2 shows structure of the atom chip.  In each layer, we have considered  $n_r$=5001 parallel long magnets of triangular cross section with magnetization 4$\pi M_z$=3.8kG and  periodicity $a$=1$\mu$m.  The bottom and top layers have thicknesses $t_1$=500nm and $t_2$=50nm, respectively.  The number of sublayers in each layer is $n$ =10 and there is no separation between them ($d$=0 in Eq.10). Each sawtooth magnetic layer can be estimated by finite number of sublayers of parallel rectangular magnetic slabs with same periodicity, width $w_i$=($i$+1)$a$/$n$ and distance $s_i$=(1-($i$+1)/$n$)$t$ from the plane $z=0$, where $i$=0, 1,..., $n$-1 is the number of the $i$th sublayer.  To create crossed permanent magnetic sawtooth layers, according to Fig.2(b), $s_i$ is subject to conditions 0$< s_i < t_1$ and $t_1< s_i < t_2$ for bottom and top arrays, respectively.\\
    For $n$=10 sublayers, great versatility can be seen in Fig.5 (a), (b) and (c) between analytical and numerical plots of magnetic field along the $x$-, $y$- and $z$-directions,  close to the center of the permanent magnetic lattice.  By adding more sublayers, we expect to have a sawtooth structure with better accuracy.  Numerical and analytical results for the 2D magnetic lattice have been compared in tables 2 and 3 for $n$= 10 and 20, respectively.  For example, symmetrical magnetic traps in $x$- and $y$- directions for $n$=10 can be obtained at $z_{min}$=1.33$\mu$m from bottom of the layers by applying the bias field $B_{1x}$= $-$5.6G and $B_{1y}$= $-$2.18G.  The minimum of the magnetic field at the center of each microtrap is 4.41G. \\
  \begin{table}
\caption{Numerical results, using Radia, and analytical ones for $^{87}$Rb in F=2
and $m_F$=2 state.  Results have been calculated for the structure shown in Fig.2.}
\label{tab:3}
\begin{tabular}{llcc}
\hline
Parameters ($\mu$k) & Definition & $n$=10 & $n$=20\\
\hline
$U_{min}$/$k_B$ &Minimum Potential &296&287  \vspace{.1cm}\\
 $\Delta U^x$/$k_B$ &Potential barrier height &201&213\\
 &in $x$-direction&&\\
  $\Delta U^y$/$k_B$ &Potential barrier height &201&213 \\
   &in $y$-direction&&\\
  $\Delta U^z$/$k_B$ &Potential barrier height &107&119\\
   &in $z$-direction&&\\
  $\hbar \omega_x$/$k_B$ &Energy level spacing in &10&10  \\
  &$x$-direction&&\\
  $\hbar \omega_y$/$k_B$ &Energy level spacing in&10&10   \\
    &$y$-direction&&\\
  $\hbar \omega_z$/$k_B$ &Energy level spacing in&15&15 \\
    &$z$-direction&&\\
\
\end{tabular}
\end{table}
  According to Fig.6, a bias magnetic field along the x-direction can change the trap frequencies.
  It is also possible to create permanent magnetic lattices with different barrier heights in the $x$- and $y$- directions.  For example, if we apply a bias magnetic field with $B_{1x}$= $-$10G and $B_{1y}$=$-$0.87G the trap depths along the $x$- and $y$-directions 4.5G and 15G, respectively.  For this bias field, corresponding frequencies along the $x$- and $y$-directions are not the same and atoms feel two different barrier heights in the $x$- and $y$-directions, respectively.  Decreasing or increasing potential barrier heights will change the rate of tunneling between lattice sites.  Different rates is useful in quantum tunnelling experiments.    \\
\begin{figure}
\begin{center}
\includegraphics[width=8.0cm]{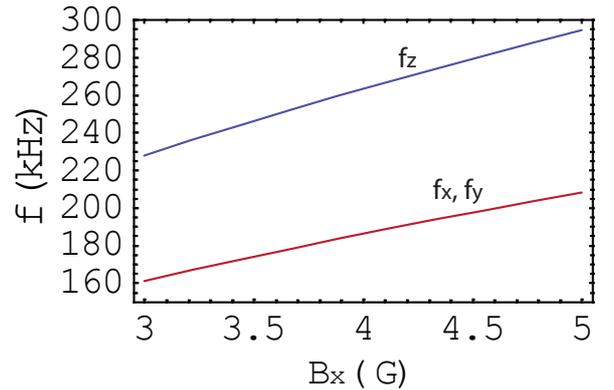}
\caption{(Color online ) Plots show effect of the ${x}$ component of the bias field on the trap frequencies for a 2D permanent magnetic lattice.} \label{fig6}
\end{center}
\end{figure}

    \section{Conclusion}
    We have introduced a new permanent magnetic lattice based on perpendicularly magnetized thin sawtooth films.  This lattice can produce Ioffe-Pritchard 2D and 3D microtraps in 1D and 2D lattices, respectively for holding ultracold atoms and BECs when a bias field is applied.  We have presented analytical expressions for parameters such as location of the microtraps, curvatures of the magnetic field and also frequencies and trap depth close to the center of the lattice. We also have shown that numerical calculations are in agreement with analytical ones for two different number of the sublayers n.  The components of homogenous external magnetic field along $x$- and $y$-directions, $B_{1x}$ and $B_{1y}$, can change the trap depth and frequencies of microtraps.   Control over the trap depths let us to handle tunneling parameters via increasing or decreasing potential barrier heights for cold atoms and BECs which has wide applications in quantum information processing and superfluid to Mott insulator quantum phase transition studies.  Trap density is $\sim$ 10$^6$ atoms/mm$^2$ for loading cold atoms and BECs.  Considering the current technology, it should be easy to manufacture the proposed nano-structure.

\begin{acknowledgement}
Amir Mohamadi would like to thank Farhood Ahani and Aref Pariz for helpful discussions.
\end{acknowledgement}


\end{document}